\documentclass[twocolumn,prl,10pt,superscriptaddress]{revtex4}
\usepackage[dvipsnames,usenames]{color}
\usepackage{graphicx}
\usepackage{dcolumn}
\usepackage{bm}
\usepackage{slashed}
\usepackage{amssymb}
\usepackage{amsmath}
\usepackage{extarrows}
\usepackage{lineno,hyperref}

\begin{document}

\title{Four-dimensional semimetals with tensor monopoles: \\ From surface states to topological responses}

\author{Yan-Qing Zhu}
\affiliation{National Laboratory of Solid State Microstructures
and School of Physics, Nanjing University, Nanjing 210093, China}
\affiliation{Center for Nonlinear Phenomena and Complex Systems, Universit$\acute{e}$ Libre de Bruxelles, CP 231, Campus Plaine, B-1050 Brussels, Belgium}

\author{Nathan Goldman}
\affiliation{Center for Nonlinear Phenomena and Complex Systems, Universit$\acute{e}$ Libre de Bruxelles, CP 231, Campus Plaine, B-1050 Brussels, Belgium}

\author{Giandomenico Palumbo}
\email{giandomenico.palumbo@gmail.com }
\affiliation{Center for Nonlinear Phenomena and Complex Systems, Universit$\acute{e}$ Libre de Bruxelles, CP 231, Campus Plaine, B-1050 Brussels, Belgium}


\date{\today}

\begin{abstract}
\noindent Quantum anomalies offer a useful guide for the exploration of transport phenomena in topological semimetals. In this work, we introduce a model describing a semimetal in four spatial dimensions, whose nodal points act like tensor monopoles in momentum space. This system is shown to exhibit monopole-to-monopole phase transitions, as signaled by a change in the value of the topological Dixmier-Douady invariant as well as by the associated surface states on its boundary. We use this model to reveal an intriguing ``4D parity magnetic effect", which stems from a parity-type anomaly. In this effect, topological currents are induced upon time-modulating the separation between the fictitious monopoles in the presence of a magnetic perturbation. Besides its theoretical implications in both condensed matter  and quantum field theory, the peculiar 4D magnetic effect revealed by our model could be measured by simulating higher-dimensional semimetals in synthetic matter.

\end{abstract}
\maketitle

\emph{Introduction.---}
Quantum anomalies play a central role in our understanding and applications of quantum field theories \cite{Bertlmann,Landsteiner}. Given a classical action, a local symmetry is ``anomalous" if it represents an obstruction to quantize the classical field theory (i.e.~the corresponding path integral cannot be made invariant with respect to both gauge symmetry and the anomalous symmetry). In high-energy physics, this obstruction prevents the existence of quantum field theories with certain symmetries, which can be cured by introducing anomaly-cancellation effects into the description~\cite{Montero}.
Such anomalous situations can however give rise to observable phenomena, such as the chiral magnetic effect related to the so-called chiral anomaly, as originally shown in Ref.~\cite{Fukushima} in the context of the quark-gluon plasma.

Quantum anomalies are not restricted to Lorentz-invariant systems. In particular, they also give rise to novel quantum effects in condensed matter physics  \cite{Yamamoto,Chernodub,Parrikar,Fujimoto,Shen,Nissinen,Stone}. For instance, it was shown that the chiral anomaly emerges in Weyl semimetals set out of equilibrium~\cite{Burkov,Grushin,Franz,Pikulin,Grushin4, Armitage}.
In this context, the corresponding chiral magnetic effect gives rise to quantized electric currents upon applying an external magnetic field.
Another well studied anomaly is the parity anomaly \cite{Redlich,Semenoff,Haldane,Witten,Lapa,Hank,Hank2,Kurkov}, which gives rise to topological effects in two-dimensional gapless (Dirac) phases \cite{Zhu2,Hughes} related to the emergence of Chern-Simons theories.

Recently, parity and chiral anomalies have been explored in multi-band models, in two and three dimensions, respectively \cite{Petkou,Burrello,Surowska}. These systems support higher-spin quasiparticles, where the Weyl-like cones become multi-fold degenerate \cite{Bercioux,Lan,Kennett,Wieder,Juricic,Bradlyn,Burrello2,Grushin2,Grushin3,Ezawa2016}.
Interestingly, the parity anomaly can also exist in four and six dimensions~\cite{Moore,Grimm}, which suggests relevant implications in the context of higher-dimensional synthetic topological matter~\cite{Goldman2,Zilberberg,Lee}.
Besides, it has been shown that a three-band model in four dimensions can give rise to a gapless topological phase characterized by a tensor monopole~\cite{Palumbo2018,Palumbo2019}, whose topological nature is established by the Dixmier-Douady ($\mathcal{DD}$) invariant~\cite{Mathai,Murray}. Their topological response in the presence of an electromagnetic field has remained unexplored.


The goal of this work is two-fold. First, we introduce a four dimensional lattice model that supports both gapless spin-1/2 and spin-3/2  birefringent fermions, depending on symmetries. When both CP (combined charge conjugation $C$ and inversion $P$ symmetries) and chiral symmetries are preserved, the Dirac cones are associated with $\mathbb{Z}_{2}$ monopoles in momentum space~\cite{YXZhao2016,YXZhao2017}. By breaking CP while preserving sublattice (chiral) symmetry, the topological semimetal phase is instead characterized by the $\mathcal{DD}$ invariant in the bulk, which one can associate to fictitious tensor monopoles~\cite{Palumbo2018,Palumbo2019}. This represents a monopole-to-monopole topological phase transition. We show the presence of topologically-protected Fermi arcs on the three-dimensional boundary, and describe their modification across the transition.
Secondly, we identify a novel quantum effect, coined ``parity magnetic effect", which arises in presence of a magnetic field and can be attributed to a parity anomaly. This is a peculiar topological effect associated to the existence of quantized topological currents in 4D semimetals. This effect could be measured in quantum-engineered settings using a synthetic dimension \cite{HMPrice2020,Ozawa2016,Ozawa2019}.

\begin{figure}[htbp]\centering
\includegraphics[width=8.5cm]{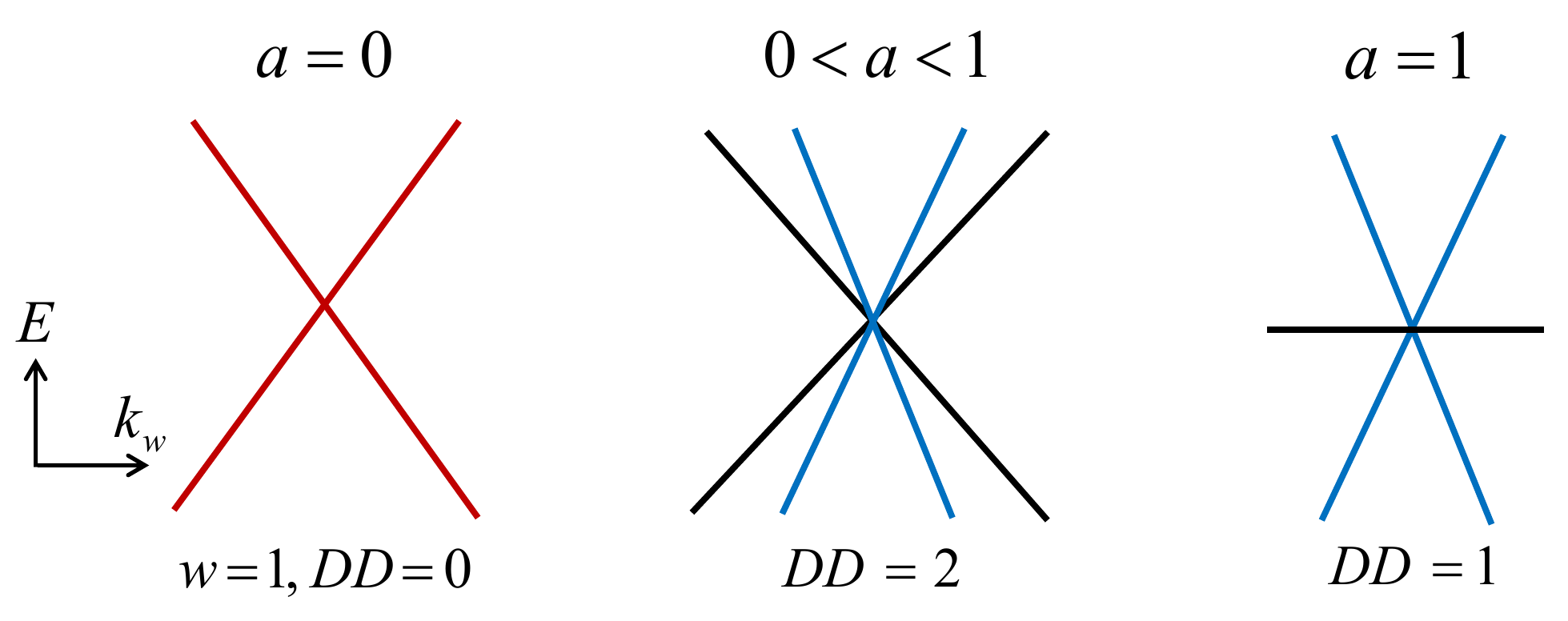}
 \caption{(Color online) Monopole-to-monopole transition:~Schematic spectra $E(k_w)$ of $H_+$ at $k_{x,y,z}=0$. When $a=0$, the two-fold degenerate spectrum (red) hosts a $\mathbb{Z}_2$ monopole with winding number $w=1$. For $a\neq 0,\pm1$, the degeneracy is lifted and the nodal point hosts a tensor monopole captured by a non-zero $\mathcal{DD}$ invariant:~$\mathcal{DD}\!=\!2$. When $a=\pm 1$, the two middle bands become perfectly flat (black) and the low-energy band (blue) contributes to $\mathcal{DD}\!=\!1$. } \label{Spectra}
\end{figure}

\emph{4D topological semimetals.---}
We start by considering a four-band Hamiltonian for spinless fermions on a four-dimensional lattice. The corresponding momentum-space Hamiltonian is given by
\begin{equation}
\begin{aligned}\label{HamMM}
H(\boldsymbol k)&=d_x\tilde{\Gamma}_x+d_y\tilde{\Gamma}_y+ d_z\tilde{\Gamma}_z+d_w\tilde{\Gamma}_w,
\end{aligned}
\end{equation}
with the four-component Bloch vector defined as
\begin{equation}
\begin{aligned}
d_x=2J\sin k_x,~d_y=2J\sin k_y,~d_z=2J\sin k_z,\\
d_w=2J(M-\cos k_x-\cos k_y-\cos k_z-\cos k_w).
\end{aligned}
\end{equation}
Here $J$ is the hopping amplitude on the 4D lattice, $M$ is a tunable parameter and the $4\times4$ matrices $\tilde{\Gamma}_i$
read
\begin{equation}
\begin{aligned}
\tilde{\Gamma}_x=\sigma_0\otimes\sigma_1+a\sigma_1\otimes\sigma_0,~~
\tilde{\Gamma}_y=\sigma_2\otimes\sigma_3+a\sigma_3\otimes\sigma_2,\\
\tilde{\Gamma}_z=\sigma_0\otimes\sigma_2+a\sigma_2\otimes\sigma_0,~~
\tilde{\Gamma}_w=\sigma_1\otimes\sigma_3+a\sigma_3\otimes\sigma_1, \nonumber \\
 \end{aligned}
\end{equation}
where $\sigma_i$ are Pauli matrices and $a$ is a constant parameter. These matrices only satisfy the Clifford algebra for $a=0$ (``Dirac regime"). When $a\ne0$, the Hamiltonian (\ref{HamMM}) supports spin-3/2 birefringent quasiparticles similarly to previous models in lower dimensions \cite{Lan,Kennett,Juricic,Bradlyn,Grushin2}.
Besides, this Hamiltonian preserves a chiral (sublattice) symmetry, $\{S,H\}=0$ with $S=\sigma_3\otimes\sigma_3$.
Its spectrum reads
\begin{equation}
E(\boldsymbol k)=\pm (1\pm a)\sqrt{d_x^2+d_y^2+d_z^2+d_w^2}.
\end{equation}
Notice that the two middle bands become perfectly flat when $a=\pm 1$ [Fig.~\ref{Spectra}].
For $2<M<4$ and $a\neq 0$, there exists a single pair of Dirac-like cones in the first Brillouin zone (BZ) separated along the $k_w$ axis and located at ${\boldsymbol K}_{\pm} =(0,0,0,\pm
\arccos k_m)$ with $k_m=M-3$. For convenience and without loss of generality, we focus on
the low-energy effective Hamiltonians near the nodal points ${\boldsymbol K}_{\pm}= (0,0,0,\pm \pi/2)$ for $M=3$,
\begin{equation}\label{Ham_Mp}
H_{\pm}({\boldsymbol q})=v q_{\pm,x}\tilde{\Gamma}_x+v q_{\pm,y}\tilde{\Gamma}_y+v q_{\pm,z}\tilde{\Gamma}_z\pm v q_{\pm,w}\tilde{\Gamma}_w,
\end{equation}
where $v=2J$ with $J>0$ and the effective momenta ${\boldsymbol q}_\pm={\boldsymbol k}-{\boldsymbol K}_{\pm}$.
When $a\!=\!0$, the Dirac cones are protected by combined CP-symmetry, $\{CP,H\}=0$, where $CP=\sigma_1\otimes\sigma_2\hat{K}$,  $\hat{K}$ is the complex conjugate, and $(CP)^2=-1$. Thus, they behave like monopoles carrying a $\mathbb{Z}_2$ charge, as studied in Ref. \cite{YXZhao2016}. For $a\neq0$, the combined CP-symmetry is broken and the Hamiltonian (\ref{HamMM}) only preserves chiral symmetry. In this regime, the system belongs to class AIII and the nodal points behave like tensor monopoles, which are characterized by a $\mathbb{Z}$ invariant \cite{Palumbo2018,Palumbo2019}. Our model thus exhibits a monopole-to-monopole phase transition upon tuning $a$; see Fig.~(\ref{Spectra}). In our $4\!\times\!4$ representation, there only exists a single mass term (proportional to $\sigma_3 \otimes \sigma_3$) that breaks CP and opens a bulk gap; however, this term simultaneously breaks the chiral symmetry $S$; hence, this perturbation would open a gap for both the $\mathbb{Z}_2$ and $\mathbb{Z}$ cases. The situation would be different in a $8\!\times\!8$ representation, where there exists a mass term that breaks S without breaking CP~\cite{SM}.

We now further characterize these two types of monopoles, by focusing on $H_{+}$ in Eq.~\eqref{Ham_Mp}.
Since the Hamiltonian preserves chiral (sublattice) symmetry in both regimes, one can calculate
 the winding number associated with the mapping ${\boldsymbol q}\backslash\{0\}\in \mathbb{S}^3$ $\rightarrow$ $\mathbf{d}/|\mathbf{d}|\in \mathbb{S}^3$, where the three-dimensional unit sphere $\mathbb{S}^3$ encloses the monopole in $q$-space. The corresponding winding number is given by \cite{Schnyder2008,DWZhang2018,Neupert2012}
\begin{equation}\label{WindEq}
w=\frac{1}{12\pi^2}\int_{\mathbb{S}^3}dq^{\mu}\wedge dq^{\nu}\wedge{dq^{\rho}}\epsilon^{ijkl}\frac{1}{|d|^4}d_i\partial_{\mu}d_j\partial_{\nu}d_k\partial_{\rho}d_l,
\end{equation}
with $d_i=v q_i$. Importantly, we obtain $w=1$ for both types of monopoles, which indicates that this winding number is not able to distinguish between the different topological-semimetal phases of our model. In order to solve this issue,
we employ the $\mathcal{DD}$ invariant, which is zero in the CP-symmetric ``Dirac" regime ($a\!=\!0$), while it is non-zero in class AIII~\cite{Palumbo2018,Palumbo2019}. This invariant can be expressed as
\begin{equation}\label{DDEq}
\mathcal{DD}=\frac{1}{2\pi^2}\int_{\mathbb{S}^3}dq^{\mu}\wedge dq^{\nu}\wedge{dq^{\rho}}~
 \sum_{n=1,2}\mathcal{H}^n_{\mu\nu\rho},
\end{equation}
where
\begin{equation}
 \mathcal{H}^{n}_{\mu\nu\rho}=\partial_{\mu}B^n_{\nu\rho}+\partial_{\nu}B^n_{\rho\mu}+\partial_{\rho}B^n_{\mu\nu} ,\label{Hfield}
 \end{equation}
 denotes the 3-form Berry curvature associated with the $n$-th eigenstate $|u_{n}({\boldsymbol q})\rangle$; here, only the two lowest bands ($n\!=\!1,2$) contribute to the $\mathcal{DD}$ invariant, as required by the half-filling condition. The 2-form tensor Berry connection $B^n_{\mu\nu}$ in Eq.~\eqref{Hfield} is defined as~\cite{Palumbo2019}
\begin{equation}
B^n_{\mu\nu}=\phi_n \mathcal{F}^n_{\mu\nu},~\phi_n=-\frac{i}{2} \log \prod^4_{\aleph=1} u^{\aleph}_{n},
\end{equation}
where  $\mathcal{F}^{n}_{\mu\nu}=\partial_{\mu}\mathcal{A}^n_{\nu}-\partial_{\nu}\mathcal{A}^n_{\mu}$ is the Berry curvature, $\mathcal{A}^n_{\mu}=\langle u_{n}|i\partial_{\mu}|u_{n}\rangle$ is the Berry connection ($\partial_{\mu}\equiv \partial_{q_\mu}$), and $u^{\aleph}_{n}$ denotes the components of $|u_{n}\rangle$.
We find $\mathcal{DD}\!=\!2$ for $a\neq 0,\pm 1$, noting that each of the two lowest bands contributes a charge $+1$. The monopole-to-monopole transition, which is signaled by a change in the value of the $\mathcal{DD}$ invariant, is illustrated in Fig. \ref{Spectra}. Besides, for the critical flat-band case ($a\!=\!\pm1$), a single non-degenerate low-energy band contributes, thus yielding $\mathcal{DD}\!=\!1$~\cite{Palumbo2018}. One also verifies that the monopoles described by $H_-$ carry the opposite tensor charge.

\begin{figure}[htbp]\centering
\includegraphics[width=8.7cm]{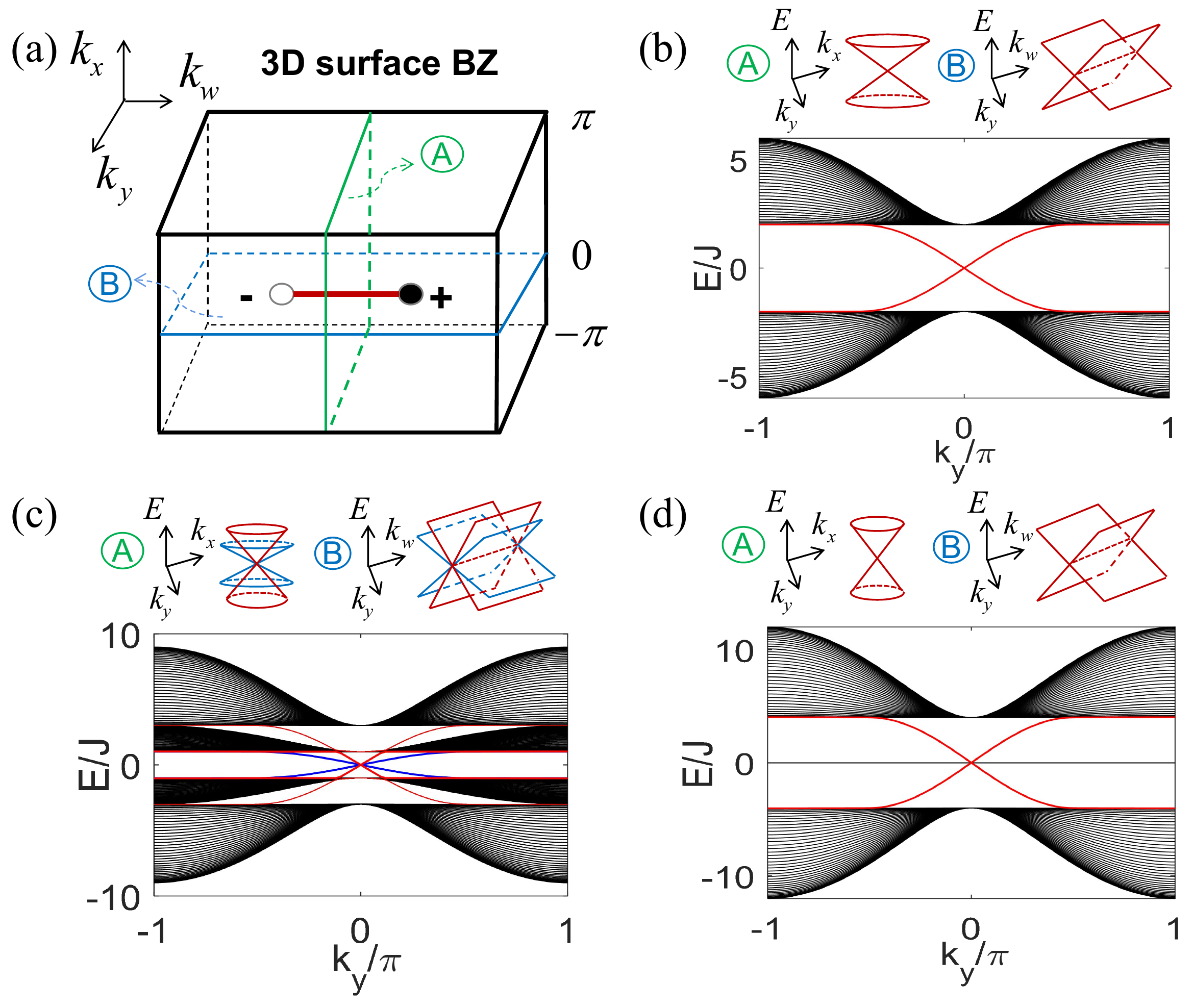}
 \caption{(Color online)  Surface spectra of the model~\eqref{HamMM} upon applying open boundary conditions along $z$. (a) Sketch of the zero-energy Fermi arc within the 3D BZ. Numerical surface spectra at $k_x=k_w=0$ are shown for (b) $a=0$, (c) $a=0.5$, and (d) $a=1$. The diagrams labeled by $A$ (resp. $B$) depict the surface spectra of the 3D topological insulator $H|_{k_w=0}$ (resp. the surface spectra of the 3D gapless semimetal $H|_{k_x=0}$). In (c) the red (resp. blue) spectra correspond to the boundary at $z=1$ (resp. $z=L_z$). Here we set $L_z\!=\!40$.} \label{FerArc}
\end{figure}

\emph{Surface states.---}To further investigate the topological properties of the semimetal Hamiltonian~\eqref{HamMM}, we now study the surface energy spectra for $M\!=\!3$, upon applying open boundary conditions along the $z$ direction. As sketched in Fig. \ref{FerArc}(a), the zero-energy surface states depict a degenerate Fermi arc (colored in red) connecting two monopoles of opposite charges.  The origin of this Fermi arc can be understood from two perspectives, as we now explain.

A first viewpoint is obtained by fixing $k_w$ and by studying the surface modes of the resulting 3D gapped subsystem. Upon taking such a slice, the 3D Hamiltonian $H|_{k_w=k_w^0}$ can either describe a $\mathbb{Z}_2$  topological insulator ($a=0$) or a chiral topological insulator ($a\neq 0$)\cite{Schnyder2008,Schnyder2009,Hosur2010,Neupert2012,STWang2014}. We find different regimes as a function of $k_w^0$:~the 3D subsystem is non-trivial [with $\mathbb{Z}_2$ index $w\!=\!1$ for $a\!=\!0$, and $\mathbb{Z}$-valued $\mathcal{DD}\!=\!2$ ($\mathcal{DD}=1$) for $a\!\neq\! 0,\pm 1$ ($a=\pm1$)], within the range $k_w^0\in (-\pi/2,\pi/2)$, and trivial otherwise. Here, the topological invariants are calculated using Eq.~\eqref{WindEq} for $a\!=\!0$ and Eq.~\eqref{DDEq} for $a\!\neq\! 0$, upon replacing $\mathbb{S}^3$ by the 3D BZ $\mathbb{T}^3$.
These 3D topological insulators host 2D Dirac boundary states, whose dispersion is defined over the $A$ plane shown in Fig.~\ref{FerArc}(a); their zero-energy nodal point forms a degenerate line along the $k_w$ axis,~i.e., a Fermi arc connecting the two monopoles.

 Another viewpoint consists in taking a slice at fixed $k_x\!=\!0$ (or $k_y\!=\!0$). The resulting subsystem $H|_{k_x=0}$ forms a gapless metallic phase, which is similar to the (real) Dirac semimetal~\cite{YXZhao2017} for $a\!=\!0$. Its dispersion, defined over the $B$ surface in Fig.~\ref{FerArc}(a), consists of two inclined planes, whose zero-energy crossing line forms a Fermi-arc connecting the two monopoles in the bulk.

We show illustrative surface spectra in Figs. \ref{FerArc}(b)--(c). For $a\!=\!0$ [Fig. \ref{FerArc}(b)], one obtains a Dirac cone over the $A$ plane and two inclined planes over the $B$ plane. Note that these spectra are 2-fold degenerate, as they describe the surface states on both boundaries (at $z\!=\!1$ and $z\!=\!L_z\!$). This degeneracy is then lifted upon increasing $a$, as shown in Fig. \ref{FerArc}(c). When reaching $a\!=\!1$ [Fig. \ref{FerArc}(d)], the surface mode at $z\!=\!L_z$ vanishes into the zero-energy flat bulk band, while a single (non-degenerate) surface mode survives at $z\!=\!1$. These surface spectra are well described by the boundary Hamiltonian~\cite{SM}
 \begin{equation}
 H^{BS}_{\pm}=\pm(1\pm a)(k_x\sigma_1-k_y\sigma_2), ~~\text{for}~k_w\in\left(-\frac{\pi}{2}, \frac{\pi}{2}\right),
 \end{equation}
which was derived from the bulk model $H({\boldsymbol k})$ for $M\!=\!3$; here $\pm$ refers to the boundaries at $z\!=\!1$ and $z\!=\!L_z$, respectively. We note that the transformations of the boundary modes reflect the monopole-to-monopole transition in the bulk [Fig.~\ref{Spectra}].

\emph{Parity magnetic effect and topological currents.---}
We now show how to derive a universal magnetic effect for our model, by calculating quantum anomalies through quantum-field-theoretical methods. In this framework, we consider the continuum limit of our 4D topological semimetal, taking into account the single pair of monopoles at ${\boldsymbol K}_{\pm}$. The resulting $8\times8$ effective Hamiltonian, defined in 4D momentum space, reads
\begin{equation}\label{ContinHam}
H_{\text{eff}}=k_{i}\tilde{G}^{i}-b_{\mu}\tilde{G}_b^{\mu},
\end{equation}
where $i=x,y,z,w$ and $\mu=t,x,y,z,w$. Here, we introduced the dipolar momentum $b_{\mu}$, which denotes the separation of the two monopoles in momentum space, with vector ${\boldsymbol b}=({\boldsymbol K}_+-{\boldsymbol K}_-)/2$, and in energy with offset $2b_t$; the $8\times8$ matrices $\tilde{G}^{i}$ and $\tilde{G}_b^{\mu}$ are defined as
\begin{eqnarray}
  \tilde{G}^{j}=\sigma_0\otimes\tilde{\Gamma}_{j}, \tilde{G}^{w}=\sigma_3\otimes\tilde{\Gamma}_{w}, \tilde{G}^{t}_b=\sigma_3\otimes I_4, \nonumber \\  \tilde{G}^{j}_b=\sigma_3\otimes\tilde{\Gamma}_{j}, \tilde{G}^{w}_b=\sigma_0\otimes\tilde{\Gamma}_{w},
  \end{eqnarray}
  where $j=x,y,z$, and $I_4$ is the $4\times4$ identity matrix.  By implementing a Legendre transformation on Eq. (\ref{ContinHam}), the action can be written in terms of a first-order Lagrangian,
\begin{equation}
S[\bar{\psi},\psi,b]=\int d^5x\,\bar{\psi}(i\tilde{\gamma}^{\mu}\partial_{\mu}-\tilde{\gamma}^{\mu}_bb_{\mu})\psi,
\end{equation}
where $\bar{\psi}=\psi^{\dagger}\tilde{\gamma}^t$, $\tilde{\gamma}^{i}=\tilde{\gamma}^t\tilde{G}^{i}$, $\tilde{\gamma}^{\mu}_b=\tilde{\gamma}^t\tilde{G}_b^{\mu}$, with $\tilde{\gamma}^t=\sigma^0\otimes S$. In order to show the topological response of the 4D semimetal to an external electromagnetic field $A_{\mu}$, we integrate out the fermion field and obtain the following effective action
\begin{equation}\label{effaction}
S_{\text{eff}}=-i\ln \det(i\tilde{\gamma}^{\mu}D_{\mu}-\tilde{\gamma}^{\mu}_bb_{\mu}),
\end{equation}
where $D_{\mu}=\partial_{\mu}-iA_{\mu}$ is the gauge covariant derivative. This effective action $S_{\text{eff}}$ with zero mass  needs to be regularized due to ultraviolet divergences \cite{Redlich}. However, the regularization explicitly breaks certain symmetries of the original action, hence giving rise to anomalies as we now show.


We use the standard Pauli-Villars method~\cite{Hank2} to obtain the topological action in the low-energy regime, which consists in introducing a mass term $\tilde{m}\bar{\psi}\psi$ with $\tilde{m}=m-\alpha k^2$. To reveal the ``parity" anomaly~\cite{FNpre}, we first consider the Dirac case ($a=0$); we determine the effective Chern-Simons action, by calculating a one-loop triangle diagram~\cite{SM,Qi}, and we obtain
\begin{equation}\label{CS}
S_{\text{top}}=\frac{C_2}{4\pi^2}\int d^5x\, \epsilon^{\mu\nu\lambda\rho\sigma} b_{\mu}\partial_{\nu}A_{\lambda}\partial_{\rho}A_{\sigma},
\end{equation}
where $C_2=-\left[\text{sgn}(m)+\text{sgn}(\alpha)\right]/2$ is nothing but the second Chern number of the gapped system described by $H_+$ with the regularized mass $\tilde{m}$~\cite{Golterman,Fukaya,Hughes}. The presence of a second Chern number in the description of a 4D semimetal is analog to the appearance of the first Chern number in 2D topological semimetals~\cite{Zhu2,Hughes}.

We find that a similar calculation~\cite{SM} can be performed for the spin-3/2 case ($a\!\ne\!0,\pm1$), yielding the same topological action in Eq.~(\ref{CS}). In that case, the anomaly takes the form of a ``4D sublattice anomaly", which shares features of the 4D parity anomaly~\cite{FN}. In the flat-band limit ($a\!=\!\pm1$), the system remains gapless in the presence of the mass $\tilde{m}$; in this case, $C_2$ diverges, and the topological action is ill-defined.

\begin{figure}[htbp]\centering
\includegraphics[width=8.8cm]{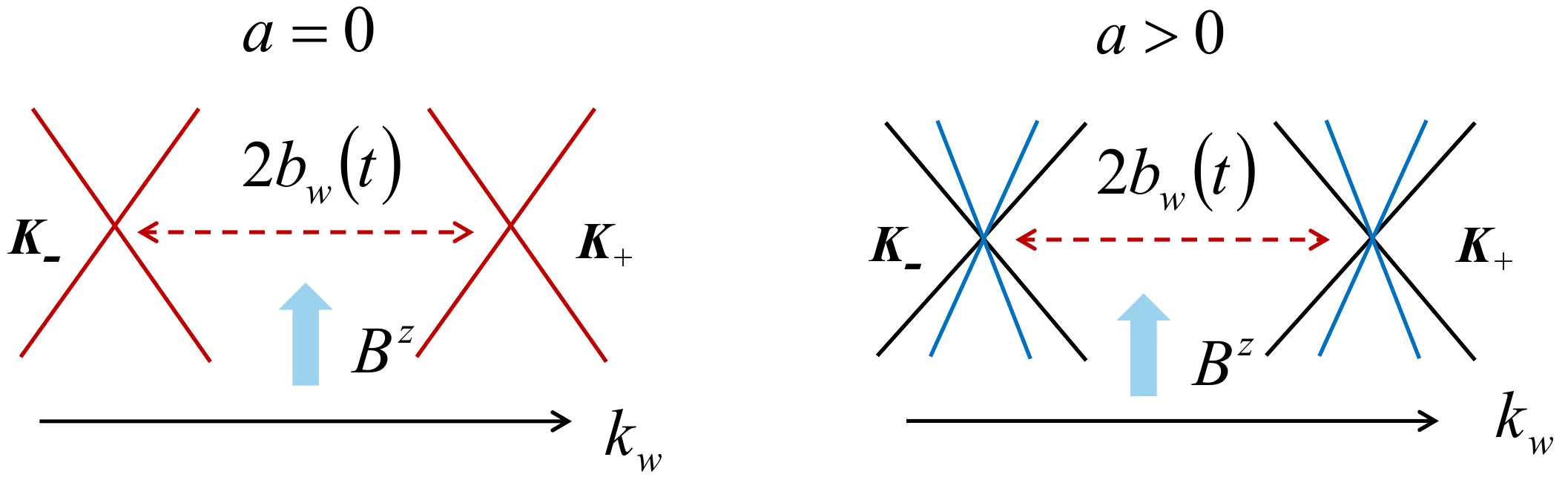}
 \caption{(Color online) The parity magnetic effect: A topological current $J^z$ can be induced
by a pair of 4D monopoles upon modulating the monopole separation $b_w(t)$ in the presence of a weak magnetic field $B^z$; see Eq.~\eqref{main_result}. } \label{Current}
\end{figure}

Based on these results, we find that the topological response current is universal for both spin-1/2 and spin-3/2  birefringent quasiparticles, and it is given by
\begin{equation}
J^{\mu}=\frac{\delta S_{\text{top}}}{\delta A_{\mu}}=\frac{C_2}{2\pi^2} \epsilon^{\mu\nu\lambda\rho\sigma} \partial_{\nu}b_{\lambda}\partial_{\rho}A_{\sigma}.\label{eq_parity_effect}
\end{equation}
This result describes the ``parity magnetic effect'' exhibited by our 4D-semimetal model, as we now further illustrate. For concreteness, let us consider the response of our system to a static and uniform magnetic field (i.e.~$A_y\!=\!xB^z$ and $A_{x,z,w,t}\!=\!0$), and to a simultaneous time-dependent modulation of the cones separation, $b_w\!=\! \arccos[M(t)-3]$; see Fig. \ref{Current}. In this case, a single component of the Faraday tensor ($F_{xy}=B^z$) contributes to Eq.~\eqref{eq_parity_effect}, which yields the topological response
\begin{equation}
J^{z}=\frac{C_2}{2\pi^2} (\partial_{t}b_{w}) B^z.\label{main_result}
\end{equation}
In this effect, the separation vector $b_{w}$ plays the role of an effective axial gauge field~\cite{Grushin2D}, whose time dependence induces an effective electric field $E_w\!=\!\partial_tb_w$.
The ``parity magnetic effect" in Eq.~\eqref{main_result} constitutes a central result of this work; it represents a unique topological response of 4D gapless topological phases, in direct analogy with the chiral magnetic effect exhibited by 3D Weyl semimetals~\cite{Burkov,Grushin,Franz,Pikulin,Grushin4, Armitage}. 

We note that the parity magnetic effect in Eq.~\eqref{main_result} could be experimental studied in 3D quantum-engineered setups extended by a synthetic dimension~\cite{Ozawa2019}, as could be realized for cold atoms in optical lattices~\cite{Goldman3}, for photons in arrays of ring resonators~\cite{Ozawa2016}, or in electric circuits~\cite{Wang2020}. The time-varying component $b_w(t)$ could be induced through a periodic driving of an on-site coupling, as proposed in Ref.~\cite{Zhu2} for a 2D-semimetal setting.

Finally, one may wonder whether the $\mathcal{DD}$ invariant, which characterizes the topology of the nodal points, also plays a role in 4D magnetic effects. In 3D Weyl semimetals, the chiral magnetic effect was shown to be directly related to the existence of Fermi arcs~\cite{Beenakker}. Similarly, we expect the Fermi arcs analysed in Fig.~\ref{FerArc} to play a similar role in 4D magnetic effects. In this framework, the $\mathcal{DD}$ invariant could constitute a key element.

\emph{Conclusions.---} We have introduced and analyzed a 4D semimetal model, whose nodal points can be associated to tensor monopoles characterized by the $\mathcal{DD}$ invariant. The system has topological Fermi arcs on its boundary, which are protected by the sublattice (chiral) symmetry. This model reveals a novel type of topological response, the parity magnetic effect, according to which a topological current can be induced by combining a magnetic perturbation with a time-modulation of the band structure. Our results expand our knowledge of quantum anomalies and their corresponding physical effects in higher-dimensional topological phases of matter, and they suggest interesting explorations in synthetic matter.\\

We thank Shi-Liang Zhu for helpful discussions. Work in Brussels is supported by the FRS-FNRS and the ERC through the Starting Grant project TopoCold. Y.~Q.~Z. also acknowledges financial support from the China Scholarship Council.

\clearpage
\newpage
\onecolumngrid
\begin{appendix}
\section*{\large{Supplementary materials}}

\section{$\mathbb{Z}_2$ Dirac and $\mathbb{Z}$ tensor monopoles in 4D}
The continuum Hamiltonian of a novel four-dimensional monopole we consider in the main text is,
\begin{equation}
\begin{aligned}\label{YMHam}
\mathcal{H}&=v_xk_x\tilde{\Gamma}_x+v_yk_y\tilde{\Gamma}_y+ v_zk_z\tilde{\Gamma}_z+v_wk_w\tilde{\Gamma}_w,
\end{aligned}
\end{equation}
with $a$-dependent $\tilde{\Gamma}_i$ are given by the main text. Notice, these matrices satisfy the Clifford algebra only for $a=0$.  Moreover, this Hamiltonian always preserve the chiral (sublattice) symmetry $\{S,\mathcal{H}\}=0$ with $S=\sigma_3\otimes\sigma_3$. For $a\neq 0$, the model belongs to class AIII. Its spectrum reads
\begin{equation}\label{Spectrum}
E(\boldsymbol k)=\pm(1\pm a) \sqrt{k_x^2+k_y^2+k_z^2+k_w^2},
\end{equation}
above we set the Fermi velocity $|v_i|=1$ with $i=x,y,z,w$ hereafter.
For $a\neq 0$, there is a band crossing at ${\boldsymbol k}=0$ which is nothing but a 4D tensor monopole. The topological charge of a tensor is characterized by the $\mathbb{Z}$-type DD invariant defined on a three-dimensional sphere $\mathbb{S}^3$ that enclosing the monopole,
\begin{equation}
\begin{aligned}
\mathcal{DD}&=\frac{1}{2\pi^2}\int_{\mathbb{S}^3}dk^{\mu}\wedge dk^{\nu}\wedge{dk^{\rho}}~
 \sum_{n\in occ.}\mathcal{H}^n_{\mu\nu\rho},\\
 &=\text{sgn}(v_xv_yv_zv_w)Q,
\end{aligned}
\end{equation}
where $Q=2$ for $a\neq 0, \pm1$, $Q=1$ for $a=\pm 1$, and $\mathcal{H}^n_{\mu\nu\rho}$ being the tensor Berry curvature as defined in the main text.

When $a=0$ , the Dirac cones are protected by combined charge conjugate $C$ and inversion $P$ symmetries  as $\{CP,\mathcal{H}\}=0$, where $CP=\sigma_1\otimes\sigma_2\hat{K}$, and $(CP)^2=-1$.  Thus, they behave like monopoles carrying a $\mathbb{Z}_2$ charge as studied in Ref. \cite{YXZhao2016}. Due to this monopole also preserves chiral symmetry, it can be characterized by the winding number,
\begin{equation}\label{WindEq}
w=\frac{1}{12\pi^2}\int_{\mathbb{S}^3}dk^{\mu}\wedge dk^{\nu}\wedge{dk^{\rho}}\epsilon^{ijkl}\frac{1}{|d|^4}d_i\partial_{\mu}d_j\partial_{\nu}d_k\partial_{\rho}d_l,
\end{equation}
with $d_i=v k_i$, and  takes the value $w=1$.

Here, we emphasize that the $\mathbb{Z}_2$ Dirac monopole of model (\ref{YMHam}) are fundamentally different from those with $\mathbb{Z}$ classification tensor monopoles in symmetry class AIII, although in both cases $\sigma_3\otimes\sigma_3$ terms are symmetry forbidden. To illustrate the $\mathbb{Z}_2$ nature of the monopole, we consider a doubled version of $\mathcal{H}$, namely,
\begin{equation}
\mathcal{H}_{double}=\sigma_0\otimes\mathcal{H},  ~S=\sigma_0\otimes\sigma_3\otimes\sigma_3, ~ CP=\sigma_0\otimes\sigma_1\otimes\sigma_2\hat{K}.
\end{equation}
It is found that there are CP-preserving perturbations (when $a=0$),
for instance, $\mathcal{H}'=m\sigma_2\otimes \sigma_3\otimes\sigma_3$, that open up a full gap. However, all of these gap opening perturbations are
forbidden by chiral symmetry because $\{S,\mathcal{H}'\}\neq0$. Hence, the discussed $\mathbb{Z}_2$ monopoles ($a=0$) are clearly distinct from the $\mathbb{Z}$ tensor monopoles ($a\neq 0$) of
class AIII.  The understanding of this results are as follows: when we consider two copy of our model $\mathcal{H}$, the charge of a monopole should be double. But for the $\mathbb{Z}_2$ monopole, the charge now is topologically equivalent to 0 due to its $\mathbb{Z}_2$ nature, because
\begin{equation}
w=1+1~\text{mod}~ 2=0, ~~ \mathbb{Z}_2+\mathbb{Z}_2=0,
\end{equation}
while the charge of tensor monopole is double,
\begin{equation}
\mathcal{DD}=2\text{sgn}(v_xv_yv_zv_w)Q,~~\mathbb{Z}+\mathbb{Z}=2\mathbb{Z},
\end{equation}
due to its $\mathbb{Z}$ nature. Therefore, the $\mathbb{Z}_2$ charge carrying zero monopole is trivial, a CP-preserving perturbation will open up a full gap while the $\mathbb{Z}$ tensor monopole carrying nontrivial $2\mathbb{Z}$ charge and thus a chiral-preserving perturbation will not destroy them but just change the position of the tensor monopole. From this point of view, we can clearly see the differences between $\mathbb{Z}_2$ Dirac and $\mathbb{Z}$ tensor monopoles described by Hamiltonian $\mathcal{H}$.  The Fermi arcs for the lattice version of these two cases can be studied similarly.

\section{Fermi arcs boundary states}
The corresponding lattice model can be obtained by letting $k_i\rightarrow  d_i$,
 then we have
\begin{equation}
\begin{aligned}\label{HamMM}
\mathcal{H}_{TSM}&=d_x\tilde{\Gamma}_x+d_y\tilde{\Gamma}_y+ d_z\tilde{\Gamma}_z+d_w\tilde{\Gamma}_w,
\end{aligned}
\end{equation}
with the four-component Bloch vector as
\begin{equation}
\begin{aligned}\label{Blochvector}
d_x&=\sin k_x,~d_y=\sin k_y,~d_z=\sin k_z,\\
d_w&=(M-\cos k_x-\cos k_y-\cos k_z-\cos k_w),
\end{aligned}
\end{equation}
with $i=x,y,z,w$, $M$ is a tunable parameter. Its spectrum has the same expression in Eq. (\ref{Spectrum}) by replacing $k_i$ with $d_i$. This model preserves CP symmetry when $a=0$, while it preserves chiral belongs to class AIII when $a\neq 0$.   For $2<M <4$ , there exists a single
pair of monopoles in the first Brillouin zone (BZ) separated along the $k_w$ axis and located at $k_w =(0, 0, 0,\pm \arccos k_m)$ with $k_m=M-3$. For convenience
and without loss of generality, we focus on the case that hosts a single pair of monopoles ${\boldsymbol K}_{\pm}=(0,0,0,\pm \pi/2)$ with $M=3$.

To study the boundary states of this situation, we expand the lattice model $\mathcal{H}_{TSM}$ around the origin $(0,0,0)$ in the 3D BZ, then we have
\begin{equation}\label{ContinM}
\mathcal{H}_{Eff}=k_x\tilde{\Gamma}_x+k_y\tilde{\Gamma}_y+k_z\tilde{\Gamma}_z+(m-\alpha k^2)\tilde{\Gamma}_w,
\end{equation}
with $m=-\cos k_w$, $k^2={k_x^2+k_y^2+k_z^2}$,  and $\alpha=-\frac{1}{2}$.
Its boundary low-energy effective theory can be formulated systematically through the perturbation theory of quantum mechanics which is a generalization of that in Ref. \cite{SQShen,YXZhao2014}. Consider that a boundary at $z=0$ is on the left of a 4D model ($k_z\rightarrow -i\partial_z$) and translation invariance is still preserved along the
other 3 directions, i.e., $k_i (i=x,y,w)$ is still a good quantum number. To implement the perturbation method,
we will first identify the gapless subspace of the model  residing
on the boundary, i.e., concentrated near $z=0$, and then
compute the transition elements in this subspace by regarding
the remaining translation invariant terms as perturbations.

For a semi-infinite chain with $z\geq 0$, we consider an open boundary condition
at $z=0$. To seek the solution of the bound state near the boundary which satisfies $\phi(z=0)=\phi(z\rightarrow+\infty)=0$, we have
\begin{equation}
\left[-i\tilde{\Gamma}_z\partial_z+(m+\alpha\partial_z^2)\tilde{\Gamma}_w\right]\phi(z)=0,
\end{equation}
where momenta along the other directions are set to be zero since only the ground state is relevant at present. The above equation can be rewritten as
\begin{equation}
\left[\partial_z+(m+\alpha\partial_z^2)i\tilde{\Gamma}_z^{-1}\tilde{\Gamma}_w\right]\phi(z)=0.
\end{equation}
Assuming $\phi(z)=\chi_{\eta}f(z)$, where $\eta=\pm1$ and $\chi_{\eta}$ is the eigenvector of $i\tilde{\Gamma}_z^{-1}\tilde{\Gamma}_w=-\sigma_1\otimes\sigma_1$ with $\eta$ being the corresponding eigenvalue. Then,
\begin{equation}
\partial_z f(z)+\eta(m+\alpha\partial_z^2)f(z)=0,
\end{equation}
with the boundary conditions:
\begin{equation}
f(0)=0,~f(z)|_{z\rightarrow +\infty}=0.
\end{equation}
Seeking solutions with the form $f\sim e^{-\lambda z}$, we have
\begin{equation}
\lambda^2-\frac{1}{\eta\alpha}\lambda+\frac{m}{\alpha}=0,
\end{equation}
with the two  roots satisfy $\lambda_1+\lambda_2=1/\eta\alpha$, and $\lambda_1\lambda_2=m/\alpha$. The boundary state requires two positive roots, which lead to the relations:  $\eta=\text{sgn}(\alpha)$, and $\text{sgn}(m)=\text{sgn}(\alpha)$. This relations further lead to
\begin{equation}
\eta=\text{sgn}(\alpha)=-1, ~~m=-\cos k_w<0\rightarrow k_{w}\in (-\frac{\pi}{2},+\frac{\pi}{2}).
\end{equation}
 It turns out that there exists two-fold degenerate solutions as
\begin{equation}
\phi_i(x)=N\chi^i_{\eta}(e^{-\lambda_1z}-e^{-\lambda_2z}),
\end{equation}
where $N$ is the normalization factor, and $i$ labels the two degenerate eigenvectors of $i\tilde{\Gamma}_z^{-1}\tilde{\Gamma}_w$ with eigenvalue $\eta=-1$.

To obtain the low-energy effective Hamiltonian on the boundary, we consider the remaining terms along the other directions of Eq. (\ref{ContinM}),
\begin{equation}
\Delta H=\sum_{i=\{x,y\}} \left(k_i\tilde{\Gamma}_i-\alpha k_i^2\tilde{\Gamma}_w\right),
\end{equation}
as perturbations. In this way, we have a three-dimensional effective model for the boundary states:
\begin{equation}
\begin{aligned}
H_{\text{BS}}^{ij}&=\langle\chi^i|\Delta H|\chi^j\rangle,\\
\end{aligned}
\end{equation}
which leads to the effective Hamiltonian expressed as
\begin{equation}\label{Top}
H_{\text{BS}}=(1+a)\left(k_x\sigma_1-k_y\sigma_2\right), ~~\text{for}~k_{w}\in (-\frac{\pi}{2},+\frac{\pi}{2}).
\end{equation}

Similarly, for a semi-infinite chain with $z\leq 0$, we consider an open boundary condition at $z=0$. To seek the solutions of the bound state near the boundary which satisfies $\phi(z=0)=\phi(z\rightarrow-\infty)=0$, we can do the same calculations according to above derivation with $\phi(z)=\chi_{\eta}f(z)$, and seek the solutions $f(z)\sim e^{\lambda z}$ for positive roots of $\lambda$.
Then we have the boundary Hamiltonian as
\begin{equation}\label{Bottom}
H_{\text{BS}}=-(1-a)\left(k_x\sigma_1-k_y\sigma_2\right), ~~\text{for}~k_{w}\in (-\frac{\pi}{2},+\frac{\pi}{2}).
\end{equation}

In the lattice model, when we consider open boundary along $z$ direction with finite length $L_z$. Eqs. (\ref{Top}) and (\ref{Bottom}) describe the boundary Hamiltonian of model (\ref{HamMM})  at the boundary $z=1$ and $z=L_z$  respectively  when $L_z$ is large enough, which perfectly match the numerical results in the main text, as shown in Fig. 2.

\section{Chern-simons action and topological response in 4D topological semimetals}
Here, we provide a discussion about the anomaly and the derivation of the Chern-Simons action for our 4D semimetal.

To derive
a non-vanishing coefficient for the topological responses in 4D, we have introduced a finite,
symmetry-breaking mass (namely, the Pauli-Villars mass regulator) that is
sent to zero at the end of the calculation. Since the
response coefficient becomes proportional only to the sign
of the symmetry-breaking mass term, it remains nonzero even
in the limit where the symmetry breaking is removed. This
effect is the manifestation of parity anomaly in 4D as well as in 2D.
Moreover, the presence of a non-zero second Chern number implies that our 4D semimetallic system with $a=0$ can be seen as the critical phase of a 4D quantum Hall state \cite{Zhang}.
As emphasized in Ref.\cite{Witten}, the word ``parity
anomaly" is misleading because is an anomaly
in time-reversal or reflection symmetry (although in Ref.\cite{Ramamurthy}, it has been shown that also inversion symmetry can be anomalous in 2D).

In fact, our classical Hamiltonian in the Dirac case, supports a reflection symmetry along the $w$ direction, namely
\begin{equation}
	U_r^{-1} H (k_x,k_y,k_z,k_w) U_r = H (k_x,k_y,k_z,-k_w),
\end{equation}
with $U_r=\sigma_2\otimes\,\sigma_0$, which is broken at quantum level by the Pauli-Villars mass regulator. This is the essence of the parity anomaly for $a=0$. The situation changes in the spin-3/2 case because the above reflection symmetry is already broken in the classical Hamiltonian for $a\neq 0$.
To seek for an anomaly here, we need to introduce a mass regulator that breaks the sublattice (chiral) symmetry. Actually, the Pauli-Villars mass regulator adopted in the previous case is the right choice because besides reflection, it breaks also the sublattice (chiral) symmetry for any value of the parameter $a$.

Next we consider the continuum limit of our 4D topological semimetal, taking into account the single pair of monopoles at ${\boldsymbol K}_{\pm}$. The resulting $8\times8$ effective Hamiltonian, defined in 4D momentum space, reads
\begin{equation}
\begin{aligned}\label{EffHam}
H_{eff}=
\left(
  \begin{array}{cc}
    b_t+({\boldsymbol k}-{\boldsymbol b})\cdot{\tilde{\boldsymbol\Gamma}} & 0 \\
    0 & -b_t+({\boldsymbol k}'+{\boldsymbol b}')\cdot{\tilde{\boldsymbol\Gamma}} \\
  \end{array}
\right)=k_{i}\tilde{G}^{i}-b_{\mu}\tilde{G}_b^{\mu},
\end{aligned}
\end{equation}
where ${\boldsymbol k}=(k_x,k_y,k_z,k_w)$, ${\boldsymbol k}'=(k_x,k_y,k_z,-k_w)$, ${\boldsymbol b}=(b_x,b_y,b_z,b_w)$, ${\boldsymbol b}'=(b_x,b_y,b_z,-b_w)$, ${\tilde{\boldsymbol \Gamma}}=(\tilde{\Gamma}_x,\tilde{\Gamma}_y,\tilde{\Gamma}_z,\tilde{\Gamma}_w)$, $i=x,y,z,w$ and $\mu=t,x,y,z,w$. The $8\times8$ matrices $\tilde{G}^{i}$ and $\tilde{G}_b^{\mu}$ are defined as
\begin{eqnarray}
  \tilde{G}^{j}=\sigma_0\otimes\tilde{\Gamma}_{j}, \tilde{G}^{w}=\sigma_3\otimes\tilde{\Gamma}_{w}, \tilde{G}^{t}_b=\sigma_3\otimes I_4,   \tilde{G}^{j}_b=\sigma_3\otimes\tilde{\Gamma}_{j}, \tilde{G}^{w}_b=\sigma_0\otimes\tilde{\Gamma}_{w},
  \end{eqnarray}
  where $j=x,y,z$, and $I_4$ is the $4\times4$ identity matrix. By implementing a Legendre transformation on Eq. (\ref{EffHam}), the action can be written in terms of a first-order Lagrangian,
\begin{equation}
S[\bar{\psi},\psi,b]=\int d^5x\,\bar{\psi}(i\tilde{\gamma}^{\mu}\partial_{\mu}-\tilde{\gamma}^{\mu}_bb_{\mu})\psi,
\end{equation}
where $\bar{\psi}=\psi^{\dagger}\tilde{\gamma}^t$, $\tilde{\gamma}^{i}=\tilde{\gamma}^t\tilde{G}^{i}$, $\tilde{\gamma}^{\mu}_b=\tilde{\gamma}^t\tilde{G}_b^{\mu}$, with $\tilde{\gamma}^t=\sigma^0\otimes S$.

The partition function with fermions coupled to both the electromagnetic potential $A_\mu$ and vector field $b_\mu$, is given by  \cite{Redlich1984}
\begin{equation}
Z[A,b]=\int D\bar{\psi}D\psi e^{iS[\bar{\psi},\psi,A,b]}\approx e^{iS_{\text{eff}}},
\end{equation}
with
\begin{equation}
S[\bar{\psi},\psi,A,b]=\int d^5x \bar{\psi}(i\slashed{\partial}+\slashed{A}-\slashed{b})\psi,
\end{equation}
where $\slashed{\partial}=\tilde{\gamma}^{\mu}\partial_{\mu}$, $\slashed{A}=\tilde{\gamma}^{\mu}A_{\mu}$, $\slashed{b}=\tilde{\gamma}^{\mu}_bb_{\mu}$.
The effective action $S_{\text{eff}}$ reads
\begin{equation}
S_{\text{eff}}=-i\ln Z=-i \ln \det(i\slashed{\partial}+\slashed{A}-\slashed{b})=-i\text{Tr} \ln(i\slashed{\partial}+\slashed{A}-\slashed{b}),
\end{equation}
where we have employed the matrix identity: $\det Q=\exp (\text{Tr} \ln Q)$. After introducing the regulator $\tilde{m}=m-\alpha k^2$, we have $S_{\text{eff}}[A,b,\tilde{m}]=-i\text{Tr} \ln(i\slashed{\partial}-\tilde{m}+\slashed{A}-\slashed{b})$. To obtain the Chern-Simons term, we expand $S_{\text{eff}}$ into the third order of the gauge field, such that the the effective action is given by
\begin{equation}
\begin{aligned}
S_{\text{eff}}[A,b,\tilde{m}]&=-i\text{Tr} \ln G_0^{-1}(1+G_0\Sigma)=-i\left(\text{Tr} \ln G_0^{-1} +\text{Tr} \ln (1+G_0\Sigma)\right)\\
&\approx -i\left(\text{Tr} \ln G_0^{-1}+\text{Tr}  (G_0\Sigma)-\frac{1}{2}\text{Tr} (G_0\Sigma G_0\Sigma)+\frac{1}{3} \text{Tr} (G_0\Sigma G_0\Sigma G_0\Sigma)+...\right),
\end{aligned}
\end{equation}
here $G_0^{-1}=i\slashed{\partial}-\tilde{m}$ is the inverse of the propagator and $\Sigma=\slashed{A}-\slashed{b}$.  Since we are seeking an induced Chern-Simons term, we focus on the following terms: $-\frac{1}{3}\text{Tr}\left(G_0\slashed{b}G_0\slashed{A}G_0\slashed{A}\right)$, $-\frac{1}{3}\text{Tr}\left(G_0\slashed{A}G_0\slashed{b}G_0\slashed{A}\right)$ and $-\frac{1}{3}\text{Tr}\left(G_0\slashed{A}G_0\slashed{A}G_0\slashed{b}\right)$, which contribute all equally to the effective topological action.  Moreover, $G_0$ and $\tilde{\gamma}^{\mu}(\tilde{\gamma}^{\mu}_b)$ are block diagonal and each $(4\times4)$ block part contributes equally. This the effective topological action reads
\begin{equation}
\begin{aligned}\label{trace}
S_{\text{top}}[A,b,\tilde{m}]&=3\times 2\times (-i)\times (-\frac{1}{3})\times\text{Tr}\left(G_+(\tilde{\gamma}^{\mu}_{b,+}b_{\mu})G_+(\tilde{\gamma}^{\nu}_{+}A_{\nu})G_+(\tilde{\gamma}^{\lambda}_{+}A_{\lambda})\right)\\
&=6\times\frac{i}{3}\times\text{Tr}\left(G_+\Sigma_{+}^bG_+\Sigma_{+}G_+\Sigma_{+}\right)\\
&=6c_{cs}\int d^5x \epsilon^{\mu\nu\lambda\rho\sigma}b_{\mu}\partial_{\nu}{A}_{\lambda}\partial_{\rho}{A}_{\sigma},
\end{aligned}
\end{equation}
where $\Sigma_+^b=b_\mu\partial G^{-1}_{+}/\partial{p_\mu}$, and $\Sigma_+=A_\mu\partial G^{-1}_{+}/\partial{p_\mu}$. Here we have used the relation $\tilde{\gamma}_{b,+}^{\mu}=\tilde{\gamma}_{+}^{\mu}=\partial G^{-1}_{+}/\partial{p_\mu}$, with the Green function of $\mathcal{H}_+$ being $G_+^{-1}=(p_0-\mathcal{H}_+)$. Notice that $\mathcal{H}_+$ now is modified as
\begin{equation}
\mathcal{H}_+=p_x\tilde{\Gamma}_x+p_y\tilde{\Gamma}_y+p_z\tilde{\Gamma}_z+p_w\tilde{\Gamma}_w+\tilde{m}\Gamma_0,
\end{equation}
where $\tilde{\Gamma}_0=\sigma_3\otimes\sigma_3$, and $\tilde{\Gamma}_i$ are given in the main text.  The result in the last line in Eq. (\ref{trace}) can be derived by using the Feyman rules\cite{Golterman,Fukaya, Ramamurthy,XLQi2008} and the corresponding coefficient is simply given by
\begin{equation}
\begin{aligned}\label{coefficient}
c_{cs}&=\frac{1}{3\times5!}\int\frac{d^5p}{(2\pi)^5}\epsilon^{\mu\nu\lambda\rho\sigma}\text{tr}\left(\mathcal{G}\frac{\partial \mathcal{G}^{-1}}{\partial p_{\mu}}\mathcal{G}\frac{\partial \mathcal{G}^{-1}}{\partial p_{\nu}}\mathcal{G}\frac{\partial \mathcal{G}^{-1}}{\partial p_{\lambda}}\mathcal{G}\frac{\partial \mathcal{G}^{-1}}{\partial p_{\rho}}\mathcal{G}\frac{\partial \mathcal{G}^{-1}}{\partial p_{\sigma}}\right)\\
&=\frac{1}{24\pi^2}\times\frac{\pi^2}{15}\times\int\frac{d^5p}{(2\pi)^5}\epsilon^{\mu\nu\lambda\rho\sigma}\text{tr}\left(\mathcal{G}\frac{\partial \mathcal{G}^{-1}}{\partial p_{\mu}}\mathcal{G}\frac{\partial \mathcal{G}^{-1}}{\partial p_{\nu}}\mathcal{G}\frac{\partial \mathcal{G}^{-1}}{\partial p_{\lambda}}\mathcal{G}\frac{\partial \mathcal{G}^{-1}}{\partial p_{\rho}}\mathcal{G}\frac{\partial \mathcal{G}^{-1}}{\partial p_{\sigma}}\right)=\frac{C_2}{24\pi^2},
\end{aligned}
\end{equation}
where $C_2$ is the second Chern number and with the imaginary Green's function being $\mathcal{G}(p)=ip_0-\mathcal{H}_+$.

Substituting Eq. (\ref{coefficient}) into Eq. (\ref{trace}), we obtain the induced Chern-Simons action
\begin{equation}
S_{\text{top}}[A,b,\tilde{m}]=\frac{C_2}{4\pi^2}\int d^5x \epsilon^{\mu\nu\lambda\rho\sigma}b_{\mu}\partial_{\nu}A_{\lambda}\partial_{\rho}A_{\sigma}.
\end{equation}
Because our model is gapless,  we have to regularize the action:
 $S_{\text{eff}}^{\text{Reg}}=S_{\text{eff}}[A,b,\tilde{m}=0]-S_{\text{eff}}[A,b,\tilde{m}\rightarrow+\infty]$ \cite{Redlich1984}. By varying the action with respect to the electromagnetic field $A_\mu$, we find
\begin{equation}
J^{\mu}=\frac{\delta}{\delta A_{\mu}}S_{\text{top}} =\frac{C_2}{2\pi^2} \epsilon^{\mu\nu\lambda\rho\sigma}\partial_{\nu}b_{\lambda}\partial_{\rho}A_{\sigma},
\end{equation}
which is the topological current associated to the parity magnetic effect.

To calculate the second Chern number $C_2$ for the gapped system $\mathcal{H}_+$, we employ the non-Abelian Berry connection and curvature \cite{XLQi2008,Price2015}, given by
\begin{equation}
\begin{aligned}
\mathcal{F}_{jk}&=\partial_{j}\mathcal{A}_{k}-\partial_{k}\mathcal{A}_{j}-i[\mathcal{A}_{j},\mathcal{A}_{k}],\\ \mathcal{A}_{j}^{mn}&=i\langle u_{m}|\frac{\partial}{\partial k_{j}}|u_{n}\rangle,
\end{aligned}
\end{equation}
with $j,k=\{k_x,k_y,k_z,k_w\}$ and $m,n=\{1,2\}$  such that
\begin{equation}
\begin{aligned}
C_2&=\frac{1}{8\pi^2}\int_{\mathbb{R}^4}\text{tr}~\mathcal{F}\wedge \mathcal{F}\\
&=\frac{1}{4\pi^2}\int_{\mathbb{R}^4}d^4k\,\text{tr}\left(\mathcal{F}_{xy}\mathcal{F}_{zw}+\mathcal{F}_{wx}\mathcal{F}_{zy}+\mathcal{F}_{zx}\mathcal{F}_{yw}\right)\\
&=-\frac{1}{2}\left[\text{sgn}(m)+\text{sgn}(\alpha)\right], ~~ a\neq \pm1,
\end{aligned}
\end{equation}
which is not well defined only when two bands become flat, namely for $a= \pm1$.

Notice that in this continuum model, the integrated manifold is $\mathbb{R}^4$ instead of $\mathbb{T}^4$. Moreover, to investigate the nontrivial topological response, we choose a suitable mass regulator with $\text{sgn}(m)=\text{sgn}(\alpha)$, such that $C_2=\pm 1$.

Finally,  the topological current $J^{\mu}$ is induced by a pair of $\mathbb{Z}_2$ monopoles due to the parity anomaly when $a=0$, while the current when $a\neq 0,\pm1$ is induced by dipolar moment originates from a pair of tensor monopoles with opposite charges.

\end{appendix}


\begin{thebibliography}{99}
\bibitem{FNpre} In the Dirac (spin-1/2) case, the ``parity" anomaly actually involves a reflection symmetry, rather than the parity symmetry (which is never anomalous in our model); see~\cite{SM}.
\bibitem{FN}
In the spin-3/2 case, reflection symmetry along the $w$ direction is already broken in the Hamiltonian.
To seek for an anomaly here, we need to introduce a mass regulator that breaks the sublattice (chiral) symmetry. Actually, the Pauli-Villars mass regulator adopted in the spin-1/2 case is the right choice because besides reflection, it breaks also the sublattice (chiral) symmetry for any value of the parameter $a$. For more details, see the Supp. Mat. \cite{SM}.
\end{thebibliography}
\end{document}